\documentclass[12pt,a4paper]{article}
\usepackage{amsmath,amssymb,graphicx}

\begin{document}

\title{$\theta$--instantons in SU(2) Higgs theory}

\author{
  F.~Bezrukov\footnote{Email: \texttt{fedor@ms2.inr.ac.ru}} \and
  D.~Levkov\footnote{Email: \texttt{levkov@ms2.inr.ac.ru}} \and
  {\small  Institute for Nuclear Research of the Russian Academy of Sciences,}\\
  {\small 60th October Anniversary Prospect 7a,}\\
  {\small Moscow, 117312, Russian Federation}\\
}
\maketitle

\begin{abstract}
We consider topology changing processes in SU(2)--Higgs theory.
In the Standard Model of particle physics they are accompanied by baryon--
and lepton--number non--conservation.
At fixed energy and multiplicity of initial state, these processes 
are described by classical 
$\theta$--instanton solutions. We describe these solutions and calculate 
the suppression
exponents for the probabilities of the topology changing transitions
at relatively low energies. 
\end{abstract}

\section{Introduction and Summary}
\label{sec:1}
Tunneling transitions between topologically distinct vacua
in the electroweak theory are accompanied by baryon and lepton number 
violation~\cite{'tHooft:1976fv}, which has important implications in particle
physics and cosmology. At zero energy, such processes are described
by instantons~\cite{Belavin:1975fg}, and their rate is suppressed by an
exponentially small factor $\exp(-2S^{(I)})\sim 10^{-170}$, where
$S^{(I)} = 8\pi^2/g^2$ is the instanton action. It was found
in Refs.~\cite{Ringwald:1990ee,Espinosa:1990qn} that at relatively low energies,
the inclusive cross sections  of the topology 
changing processes in particle collisions can be calculated by perturbative 
expansion about the instanton background, and that these cross sections 
grow rapidly with energy. In particular, the leading order instanton
cross section becomes unsuppressed at energies comparable to the 
sphaleron energy $E_\mathrm{sph}\sim 10\mathrm{TeV}$.
However, further 
studies~\cite{Khlebnikov:1991ue,Mueller:1991fa,Khoze:1991sa,Arnold:1991pe}
(see Refs.~\cite{Mattis:1992bj,Tinyakov:1993dr} for reviews) 
revealed that the actual expansion  parameter of the perturbation theory about 
the instanton is 
$(E/E_\mathrm{sph})^{2/3}$, so the most interesting region 
$E \gtrsim E_\mathrm{sph}$ is unreachable for analytic methods.
The results of the perturbative analysis at 
$E \ll E_\mathrm{sph}$ suggested the exponential form for the cross section,
\begin{equation}
\label{1-1}
\sigma_0(E)\propto \exp\left\{\frac{1}{g^2}F_0(E/E_\mathrm{sph})\right\}
\;,
\end{equation}
where $g$ is the weak coupling constant, while the suppression exponent
$F_0(E/E_\mathrm{sph})$ is negative and grows with energy.
To the leading order in perturbation theory about the instanton
{}
$$
F_0 = -16\pi^2 + 3
\left[\frac{3 g^4 E^4}{8 \pi^2 v^4}\right]^{1/3}
+ O(g^2 E^2/v^2)\;,
$$
where $v$ is the Higgs vacuum expectation value.

The exponential form~\eqref{1-1} implies that there might exist 
some semiclassical--type procedure which would allow one to calculate the
suppression exponent at all energies. The main difficulty with such a procedure
resides in the fact that the initial state of the process contains two highly
energetic particles and cannot be described semiclassically. A way out
of this difficulty was suggested in 
Ref.~\cite{Rubakov:1992ec}. The main idea is to study the 
instanton-like processes with parametrically large energy and initial-state
particle number, $E = \tilde{E}/g^2$, $N = \tilde{N}/g^2$. 
One can then justify the use of the semiclassical methods for the
semi--inclusive cross section $\sigma(E,N)$: the multiparticle topology 
changing processes in the weak coupling limit $g^2\to 0$ are described 
by $\theta$--instantons, solutions to a certain classical boundary value problem, 
and the multiparticle suppression exponent 
$F(\tilde{E}, \tilde{N})$ is calculated as a value of an appropriately
modified action functional on these 
solutions~\cite{Rubakov:1992ec}. We review the boundary value problem for the 
$\theta$--instantons in Section 2. 

It was argued in Refs.~\cite{Tinyakov:1992fn,Mueller:1993sc} that 
the two-particle suppression exponent can be found as a limit of the 
multiparticle one,
\begin{equation}
\label{1-2}
F_0(\tilde{E}) = \lim\limits_{\tilde{N}\to 0} F(\tilde{E},\tilde{N})\;.
\end{equation}
This conjecture was checked by explicit calculations in quantum mechanics 
in Ref.~\cite{Bonini:1999kj}.
To summarize, the $\theta$--instanton method allows one to find, after performing
the limiting procedure~\eqref{1-2}, the suppression exponent for the 
two-particle cross section at all energies.

It was shown in Ref.~\cite{Rubakov:1992ec} that at low 
energies, the  $\theta$--instanton solution can be approximated by a chain of
appropriately modified 
instantons and anti-instantons placed at certain positions along the Euclidean
time axis. Although this approximation is justified only at 
$E \ll E_\mathrm{sph}$, the approximate solutions give an idea of the form of 
$\theta$--instantons in the whole region $E < E_\mathrm{sph}$.
In Section 3 we investigate the properties of $\theta$--instantons 
at low energies in SU(2) gauge--Higgs theory, which is a close analog
of the Standard Model. We show that the chain instanton approximation gives the 
multiparticle suppression exponent $F(\tilde{E},\tilde{N})$, up to corrections 
of order $O(\tilde{E}^2)$. In the limit of small number of particles,
our result for $F(\tilde{E}, \tilde{N})$ coincides with the perturbative 
calculations of Refs.~\cite{Zakharov:1992dj,Porrati:1990rk,Khlebnikov:1991ue} 
for $F_0(\tilde{E})$, and thus
eq.~\eqref{1-2} is valid indeed. We find, however, that the low--$N$ expansion
of the suppression exponent contains a term $N\log N$. This term does 
not depend on energy and therefore the derivative $\partial F/\partial E$
is regular at $N=0$.

During the last decade, sophisticated numerical techniques of 
finding the $\theta$--instanton solutions at high energies were 
developed
~\cite{Kuznetsov:1996cm,Kuznetsov:1997sf,Bezrukov:2003yf,Bezrukov:2001dg}).
The approximate $\theta$--instanton solutions which we find in this paper may
serve as a cross check of numerical methods. In Section 3 we perform this 
check in the SU(2)--Higgs model and find that after extrapolating to low 
energies, the numerical data obtained in 
Refs.~\cite{Bezrukov:2001dg,Bezrukov:2003dg} agree with our analytical results.

\section{$T/\theta$ boundary value problem}
\label{sec:2}
\begin{figure}
  \centerline{\includegraphics[width=0.8\textwidth]{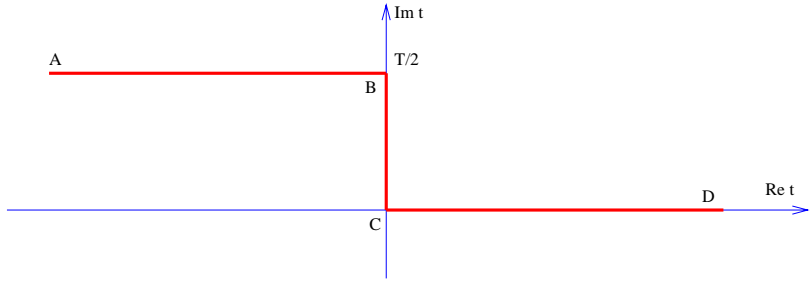}}
  \caption{Contour in the complex time plane.}
\label{fig1}
\end{figure}

The boundary value problem for the $\theta$--instanton~\cite{Rubakov:1992ec}
involves two Lagrange multipliers, $T$ and $\theta$, 
which enable one to fix 
energy and initial particle number.  The problem is naturally formulated
on the contour ABCD in complex time plane (see fig.~\ref{fig1}),
with imaginary part of the initial time equal to $T/2$.
In the internal points of the contour, the $\theta$--instanton fields, which we
denote collectively by $\varphi(\boldsymbol{x},t)$, satisfy the classical 
field equations,
\begin{subequations}
\label{2-1}
\begin{equation}
\label{2-1a}
\frac{\delta S}{\delta \varphi} = 0\;.
\end{equation}
In the asymptotic future (region D of the contour) the fields are real
and describe the evolution of the system after transition,
\begin{equation}
\label{2-1b}
\mathrm{Im}\;\varphi(\boldsymbol{x},t)\bigg\vert_\mathrm{D}  = 0\;.
\end{equation}
In the asymptotic past (part A of the contour) the fields have to be in linear 
regime and satisfy the linearized field equations. Thus
\begin{equation*}
\varphi(\boldsymbol{x},t)\bigg\vert_\mathrm{A} = \frac{1}{(2\pi)^{3/2}}\int 
\frac{d\boldsymbol{k}}{\sqrt{2 \omega_{\boldsymbol{k}}}} \left[
f_{\boldsymbol{k}} \mathrm{e}^{ - i \omega_{\boldsymbol{k}}(t - iT/2) + 
i\boldsymbol{kx}} + g_{\boldsymbol{k}}^* 
\mathrm{e}^{i \omega_{\boldsymbol{k}}(t - iT/2) - i\boldsymbol{kx}}\right]\;.
\end{equation*}
The boundary conditions in the asymptotic past relate the positive and 
negative frequency components of the solution,
\begin{equation}
\label{2-1c}
f_{\boldsymbol{k}} = \mathrm{e}^{-\theta} g_{\boldsymbol{k}}\;.
\end{equation}
\end{subequations}
Note that at $\theta\to +\infty$, the boundary conditions~\eqref{2-1c} transform 
into the Feynman ones, and the corresponding $\theta$--instanton
solution describes a transition from a state with vanishingly small number of 
particles. At finite $\theta$, equation~\eqref{2-1c} may be viewed as a 
deformation of the Feynman boundary conditions.
Note that eq.~\eqref{2-1c} implies that the $\theta$--instanton solution
is {\it necessarily complex}, except for the special periodic 
instanton~\cite{Khlebnikov:1991th} case $\theta = 0$.

By solving eqs.~\eqref{2-1a}--~\eqref{2-1c}, one finds the $\theta$--instanton solution for given values of 
the Lagrange multipliers $T$ and  $\theta$. The suppression exponent is the 
value of the following functional evaluated on this solution,
\begin{equation}
\label{2-2}
\frac{1}{g^2}F(E,N) = E T + N\theta - 2\mathrm{Im}\; S_\mathrm{ABCD}\;,
\end{equation}
where $S_\mathrm{ABCD}$ is the action calculated along the contour ABCD.
The extremization of the suppression exponent with respect to the Lagrange 
multipliers gives equations determining the values of $T$ and $\theta$:
\begin{subequations}
\begin{eqnarray}
\label{2-3a}
E = 2 \frac{\partial}{\partial T} \mathrm{Im}\; S_\mathrm{ABCD}
\;,\\
\label{2-3b}
N = 2\frac{\partial}{\partial \theta} \mathrm{Im}\; S_\mathrm{ABCD}
\;.
\end{eqnarray}
\end{subequations}
One can show that the values of $E$ and $N$ are equal to the classical energy 
and initial particle number calculated on the 
$\theta$--instanton solution,
\begin{eqnarray}
\nonumber
&& E = \int d\boldsymbol{k} \;\omega_{\boldsymbol{k}} f_{\boldsymbol{k}} 
g_{\boldsymbol{k}}\;,\\
\nonumber
&& N = \int d\boldsymbol{k} \;f_{\boldsymbol{k}} g_{\boldsymbol{k}}\;.
\end{eqnarray}

\section{$\theta$--instanton at low energy}
\label{sec:3}
\begin{figure}
  \centerline{\includegraphics[width=0.8\textwidth]{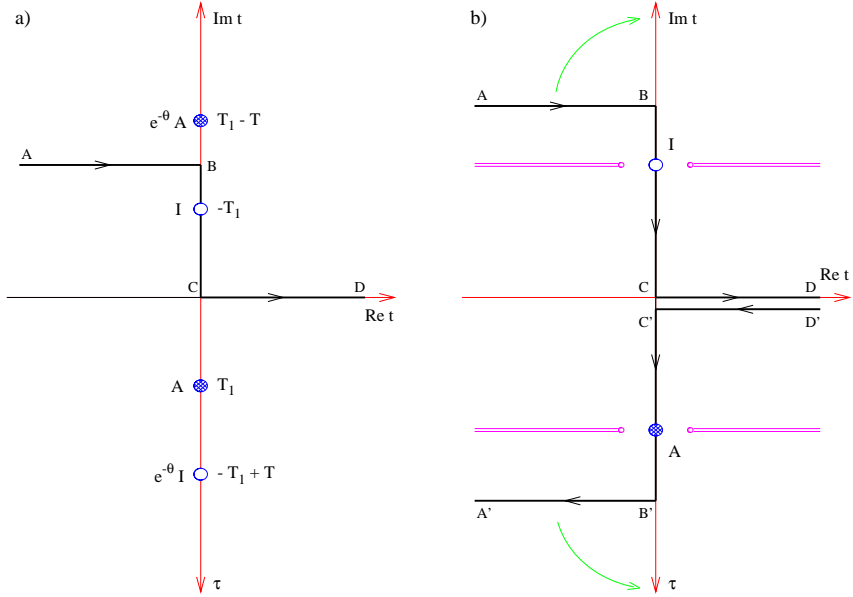}}
  \caption{a) $\theta$--instanton at low energy, 
b) instanton---anti--instanton pair.}
\label{fig2}
\end{figure}

We consider SU(2) gauge theory with a doublet Higgs field, which 
coincides with the bosonic sector of the standard electroweak 
theory with the weak mixing angle set equal to zero. 
The model possesses constrained instantons and anti--instantons
of size $\rho$, which in a singular gauge have the following 
form~\cite{Belavin:1975fg,Affleck:1981mp}:
\begin{eqnarray}
\nonumber
A_\mu^{a(I)} = \frac{2 \rho^2}{g}
\frac{\bar{\eta}^a_{\mu\nu}x_\nu}{x^2(x^2+\rho^2)}\;,\\
\nonumber
A_\mu^{a(A)} = \frac{2 \rho^2}{g}
\frac{\eta^a_{\mu\nu}x_\nu}{x^2(x^2+\rho^2)}\;,
\end{eqnarray}
where $x^0 \equiv \tau = it$ is the Euclidean time.
We construct the $\theta$--instanton 
solution by placing instantons and anti--instantons along the Euclidean time axis, 
as shown in figure~\ref{fig2}a, and suppressing them by factors 
$\mathrm{e}^{-\theta |n|}$:
\begin{equation}
\label{3-1}
A_\mu^{a(\theta)} (\boldsymbol{x}, \tau) = \sum\limits_{n=-\infty}^{+\infty}
\mathrm{e}^{-\theta|n|}\left[A_\mu^{a(I)}(\boldsymbol{x},\tau + T_1 + n T)
+ A_\mu^{a(A)}(\boldsymbol{x},\tau - T_1 + n T)\right]\;.
\end{equation}
Hence, the (suppressed) instantons sit at $\mathrm{Im}\; t = -\tau = T_1 + nT$,
while anti--instantons are placed at  $\mathrm{Im}\; t = -\tau = -T_1 + nT$,
$n = 0,\;\pm 1,\;\pm 2 \dots$.
We will see that at low energies the instanton size is small compared to the 
instanton separation, $\rho \ll T$, $\rho \ll T_1$, and $T,T_1 \ll 1/(gv)$. 
This is the basic reason 
for approximating the solution as a sum of the instanton and anti--instanton 
fields.

Let us show that the solution~\eqref{3-1} satisfies the field 
equations~\eqref{2-1a} along the contour ABCD and boundary conditions~\eqref{2-1c}
and~\eqref{2-1b} in the asymptotic regions A and D, respectively.
It is convenient to work with the Fourier transform of the 
instanton field, 
$$
A_\mu^{a(I)}(\boldsymbol{k},\tau) = \int \frac{d\boldsymbol{x}}{(2\pi)^{3/2}}
\mathrm{e}^{i\boldsymbol{kx}} A_\mu^{a(I)}(\boldsymbol{x},\tau)\;,
$$ 
with
\begin{subequations}
\begin{eqnarray}
\label{3-2a}
&&A_0^{a(I)}(\boldsymbol{k}, \tau) = - \frac{2i\rho^2}{g}\frac{\partial}
{\partial k_a} \Phi(k, \tau)\;,\\
\label{3-2b}
&&A_i^{a(I)}(\boldsymbol{k},\tau ) = -\frac{2\rho^2}{g}
\left(\delta_{ia}\tau + i \epsilon_{ija}\frac{\partial}{\partial k_j}\right)
 \Phi(k, \tau)\;,
\end{eqnarray}
where $k = |\boldsymbol{k}|$ and 
\begin{equation}
\label{3-2c}
\Phi(\boldsymbol{k},\tau) = \frac{\sqrt{2 \pi}}{4|\tau|}
\mathrm{e}^{-k|\tau|} + O(\rho^2/\tau^3)\;.
\end{equation} 
\end{subequations}
Along the contour ABCD of fig.~\ref{fig1}, the  approximate 
solution~\eqref{3-1} has the form of the superposition of 
the ``main instanton'' at $\mathrm{Im}\; t = -\tau = T_1$, and small linearized asymptotics of 
other instantons and anti--instantons. Since outside the instanton core, the
instanton field has low momenta, $k\lesssim 1/T$ 
(see eq.~\eqref{3-2c}), its interaction with the core of another instanton
is suppressed by powers of $\rho^2/T^2$. Therefore, up to 
corrections of order $O(\rho^2/T^2)$ the linear combination~\eqref{3-1} 
satisfies the field equations~\eqref{2-1a}. 

In the asymptotic regions A and D one is able to sum up the contributions
from all instantons and anti--instantons. After performing 
the transformation to the gauge $A_0^a=0$, we obtain:
\begin{subequations}
\begin{eqnarray}
\nonumber
A_i^{a(\theta)}(\boldsymbol{k},t)\bigg\vert_A &=& \frac{\rho^2\sqrt{2\pi}}
{g[1 - \mathrm{e}^{-kT-\theta}]}
\left[\mathrm{e}^{ikt} + \mathrm{e}^{-ikt-kT-\theta} \right]\;\times\\
\label{3-3a}
&&\left\{\sinh(kT_1)(\delta_{ia} - 
\frac{k_ik_a}{k^2}) + i\epsilon_{ija}\frac{k_j}{k}\cosh(kT_1)\right\}\;,\\
\nonumber
A_i^{a(\theta)}(\boldsymbol{k},t)\bigg\vert_D &=& -\frac{\rho^2\sqrt{2\pi}
\cos(kt)}{g[1 - \mathrm{e}^{-kT-\theta}]}\;\times\left\{(\delta_{ia} - 
\frac{k_ik_a}{k^2})\left[\mathrm{e}^{-kT_1} - \mathrm{e}^{kT_1-kT-\theta}
\right] - \right.\\
\label{3-3b}&&\;\;\;\;\left.
i\epsilon_{ija}\frac{k_j}{k}\left[\mathrm{e}^{-kT_1} +
\mathrm{e}^{kT_1-kT-\theta}\right]\right\}\;.
\end{eqnarray}
\end{subequations}
We see that the boundary conditions~\eqref{2-1b} and~\eqref{2-1c} are 
satisfied indeed.

To calculate the suppression exponent~\eqref{2-2} on the approximate 
solution~\eqref{3-1}, it is instructive to consider first the instanton---
anti--instanton configuration and evaluate the imaginary part of its action
along the contour ABCD (see fig.~\ref{fig2}b). Since on this contour,
 anti--instanton is complex conjugate to instanton, 
$A^{(A)}(\boldsymbol{x},t) = [A^{(I)}(\boldsymbol{x},t)]^*$, instanton---
anti--instanton configuration is $C$--symmetric, and 
we have
\begin{equation}
\label{3-4}
2\mathrm{Im}\; S_{\mathrm{ABCD}}^{(IA)} = S_{\mathrm{ABCD}}^{(IA)} - 
(S_{\mathrm{ABCD}}^{(IA)})^* = S_{\mathrm{ABCD}}^{(IA)} + 
S_{\mathrm{D'C'B'A'}}^{(IA)} = S_E^{(IA)}\;,
\end{equation}
where $S_E^{(IA)}$ denotes the Euclidean action of the instanton--- 
anti--instanton pair. Note that the (anti)instanton singularities marked in 
fig.~\ref{fig2}b by dashed lines do not permit one to move the contour ABCD
to the regions $\tau\to\pm\infty$, where (anti)instanton field is equal to
zero.
The quantity $S_E^{(IA)}$ naturally divides into the sum of the instanton and 
anti--instanton actions,
\begin{equation}
\label{3-5}
S^{(I)} = S^{(A)} = \frac{8\pi^2}{g^2} + \pi^2\rho^2v^2
\;,
\end{equation}
where the second term represents the Higgs field contribution,
and the interaction action, which was calculated 
in Refs.~\cite{Forster:1977ip,Callan:1978gz,Diakonov:1984hh},
\begin{equation}
\label{3-6}
S_\mathrm{int}^{(IA)} = - \frac{96\pi^2\rho^4}{g^2 l^4}\;.
\end{equation}
Here $l=2T_1$ is the distance between the instanton and anti--instanton, and 
corrections involving powers of $\rho^2/l^2$ in~\eqref{3-5} 
and~\eqref{3-6} are omitted. 

Finally, collecting formulas~\eqref{3-5} and~\eqref{3-6}, we have
\begin{equation}
\label{3-7}
2\mathrm{Im}\; S_\mathrm{ABCD}^{(IA)} = \frac{16\pi^2}{g^2} + 2\pi^2 \rho^2v^2 
- \frac{96\pi^2\rho^4}{l^4} + O(\rho^6/l^6)\;.
\end{equation}

The action of the $\theta$--instanton is the sum of the ``main'' instanton
action~\eqref{3-5} and the interaction terms, which, up to corrections
of order $O(\rho^6/T^6)$, are quadratic with respect to the (anti)instanton
fields. The instanton---instanton interaction is of order $O(\rho^6/l^6)$ (see  
Refs.~\cite{Forster:1977ip,Callan:1978gz,Diakonov:1984hh})
and thus does not contribute into the action to the order we study. 
Thus, the interaction action of the $\theta$--instanton is the sum of
interactions of different instanton---anti--instanton pairs.
It is clear that if both instanton and anti--instanton are situated above
(or below) the main instanton, the contour ABCD may be moved to the region
$\tau\to\pm\infty$ without crossing the singularities, and therefore the 
interaction action of such pair is equal to zero. The above argument with 
$C$--conjugation  shows that even if the instanton 
and anti--instanton are situated at different sides of the main instanton,
their interaction action equals zero. Therefore, the only non-zero terms
in the action emerge from the interaction of the ``main'' instanton with different
anti--instantons. These terms were calculated above.
Summing up all of them, we obtain
\begin{eqnarray}
\nonumber
S_\mathrm{int}^{(\theta)} &=& - \frac{96 \pi^ 2 \rho^4}{g^2}
\sum\limits_{n = -\infty}^{+\infty}
\frac{\mathrm{e}^{-\theta |n|}}{(2T_1 + nT)^4} = \\
\label{3-8}
&&-\frac{16\pi^2\rho^4}{g^2}\int\limits_0^\infty 
\frac{dk\; k^3 }{1 - \mathrm{e}^{-kT-\theta}}\left[\mathrm{e}^{-2kT_1}+
\mathrm{e}^{2kT_1 - kT-\theta}\right]\;,
\end{eqnarray}
where we have used the integral representation for the sum in last equality.
Finally, eq.~\eqref{2-2} gives the expression for the suppression exponent,
\begin{equation}
\label{3-9}
\frac{1}{g^2}F = ET + N\theta - \frac{16\pi^2}{g^2} -
2\pi^2 v^2\rho^2 + \frac{16\pi^2\rho^4}{g^2} \int\limits_0^\infty \frac{dk \;k^3}
{1 - \mathrm{e}^{-kT-\theta}}\left[\mathrm{e}^{-2kT_1}+ 
\mathrm{e}^{2kT_1 - kT - \theta}\right]
\end{equation}

\begin{figure}
  \centerline{\includegraphics[width=0.8\textwidth]{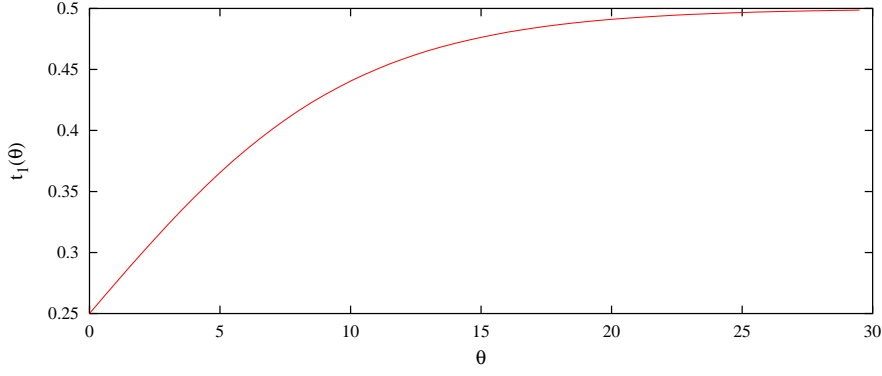}}
  \caption{The ratio $t_1 = T_1/T$ as function of $\theta$.}
\label{fig3}
\end{figure}

Note that, apart from the Lagrange multipliers $T$ and $\theta$, which are 
determined by equations~\eqref{2-3a} and~\eqref{2-3b}, the solution~\eqref{3-1}
has two free parameters: the instanton size $\rho$ and the position of the ``main''
instanton $T_1$. The values of these parameters are to be chosen to give
the extremum of the suppression exponent $F$.
The extremization of~\eqref{3-9} with respect to $T_1$ determines the ratio
$t_1\equiv T_1/T$ as a function of $\theta$. This ratio satisfies the equation
\begin{equation}
\label{3-10}
\int\limits_0^\infty \frac{dq\;q^4}{1 - \mathrm{e}^{-q-\theta}}\left[
\mathrm{e}^{2qt_1 - q - \theta} - \mathrm{e}^{-2qt_1}\right] = 0\;.
\end{equation}
When $\theta = 0$ (periodic instanton case), one finds $t_1 = 1/4$, 
so the  anti--instantons are situated exactly in the middle between the instantons.
In the limiting case ${\theta\to +\infty}$ 
(corresponding to the limit $N\to 0$), one has
$t_1\to 1/2$, and instantons approach the neighbouring anti--instantons.
As $\theta$ changes from $0$ to $+\infty$, $t_1$ smoothly interpolates
between the two limiting values, see fig.~\ref{fig3}.

It is convenient to express the saddle point values of the quantities $\rho$,
$T$ and $\theta$ in terms of two integrals:
\begin{subequations}
\begin{eqnarray}
\label{3-11a}
I_E(\theta) = \frac{1}{2}\int\limits_0^\infty \frac{dq\; q^4}
{(1 - \mathrm{e}^{-q-\theta})^2}\mathrm{e}^{-q-\theta}\cosh[2qt_1(\theta)]\;,\\
\label{3-11b}
I_N(\theta) = \frac{1}{2}\int\limits_0^\infty \frac{dq\; q^3}
{(1 - \mathrm{e}^{-q-\theta})^2}\mathrm{e}^{-q-\theta}\cosh[2qt_1(\theta)]\;.
\end{eqnarray}
\end{subequations}
The extremization of~\eqref{3-9} with respect to $\rho$ gives
\begin{equation}
\label{3-12}
\rho^2 = \frac{g^2 v^2T^4}{16 I_E(\theta)}\;,
\end{equation}
while equations~\eqref{2-3a} and~\eqref{2-3b} imply,
\begin{eqnarray}
\label{3-13}
&&T = \left[\frac{4 E I_E(\theta)}{\pi^2 g^2 v^4}\right]^{1/3}\;,\\
\label{3-14}
&&\frac{N}{E^{4/3}} = 
I_N(\theta)\left[\frac{2 }{\pi g v^2 I_E(\theta)}
\right]^{2/3}\;.
\end{eqnarray}
Equation~\eqref{3-14} determines $\theta$ explicitly as function of 
$N/E^{4/3}$. In terms of the 
integrals $I_E(\theta)$ and $I_N(\theta)$, the suppression exponent takes the form
\begin{equation}
\label{3-15}
\frac{1}{g^2}F(\theta,E) = -\frac{16\pi^2}{g^2} + 
\left[\frac{E^4}{16 \pi^2 g^2 v^4}\right]^{1/3}
\left(3I_E(\theta)^{1/3}+ 4\theta\frac{I_N(\theta)}{I_E(\theta)^{2/3}}\right)
+ O(\rho^6/g^2 T^6)
\end{equation}
Several remarks are in order:

(i) Note that the corrections to the suppression exponent calculated within 
the instanton chain approximation are 
of order $O(\rho^6/T^6)$, where 
\begin{equation}
\label{3-19}
\frac{\rho^6}{T^6} = \frac{E^2 g^2}{256 \pi^4 v^2 I_E(\theta)}\;.
\end{equation}
As $I_E(\theta)$ is a bounded function, one has 
$(\rho/T)^6 \sim O(E^2 g^2/v^2)$. Thus, our approximation is indeed valid
at $E^2 \ll v^2/g^2$. We note also that corrections due to the Higgs field 
interactions are also of order $(\rho/T)^6$.

\begin{figure}
  \centerline{\includegraphics[width=0.8\textwidth]{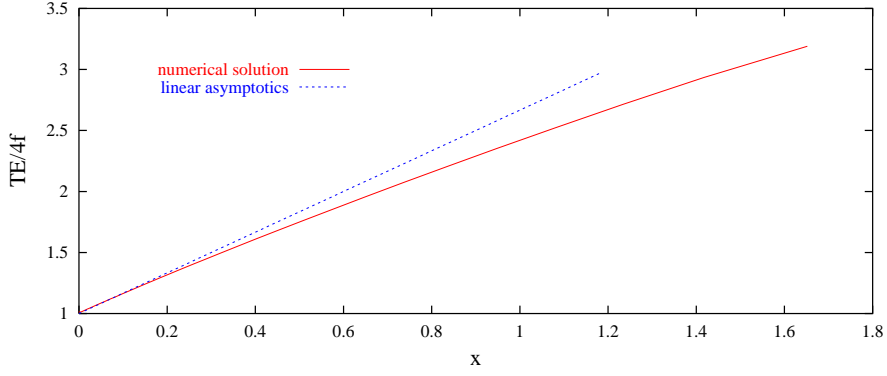}}
  \caption{Rescaled period $T$ as function of $x$ and its linear asymptotics 
at small $x$.}
\label{fig4}
\end{figure}

(ii) Although we cannot solve eq.~\eqref{3-14} analytically
in the entire region
$\theta\in [0;\;+\infty)$, we are able to find the asymptotics of large
$\theta$ (small $N$). In this way we obtain at small $N$,
\begin{equation}
\label{3-16}
\frac{1}{g^2}F = -\frac{16\pi^2}{g^2} + f\left\{
3-5x\log x +5x+\frac53 x^2 - \frac{35}{27} x^3 + O(x^4)\right\}\;,
\end{equation} 
where 
\begin{equation}
\label{3-17}
f \equiv \left[\frac{3E^4}{8\pi^2 g^2 v^4}\right]^{1/3}\;,
\end{equation}
and
\begin{equation}
\label{3-23,5}
x\equiv N/f\;.
\end{equation}
Note that although the limit $N\to 0$ is singular,
$N\log N$ term does not depend on energy (at least,
to the first order of the perturbation theory). Thus, the ``period'' 
$T(E,N)$ is a regular function at $N=0$:
\begin{equation}
\label{3-18}
T = \frac{4 f}{E}\left\{1+ \frac53 x - \frac59 x^2 +
\frac{70}{81} x^3 + O(x^4)\right\}\;.
\end{equation}
This validates the method of Ref.~\cite{Bezrukov:2003dg}, where the 
period is extrapolated to the region of small $N$ by polynomials.
Rescaled period $T$ as function of $x=N/f$ at fixed energy, together with
its linear asymptotics at small $N$ is shown in fig.~\ref{fig4}.

(iii) In the limiting case $N=0$ (two--particle collisions)
and $\theta=0$ (periodic instanton) the leading terms of order $(g^2E)^{4/3}$ 
in the suppression exponent were found in Refs.~\cite{Khlebnikov:1991ue} 
and~\cite{Khlebnikov:1991th}, respectively.
From eqs.~\eqref{3-15},~\eqref{3-16} we obtain
\begin{eqnarray}
\nonumber
&&\frac{1}{g^2}F\bigg\vert_{\theta=0} = -\frac{16\pi^2}{g^2} + 
\frac32\left[\frac{\pi^2E^4}{g^2 v^4}\right]^{1/3} + O(E^2/v^2)\;,\\
\nonumber
&&\frac{1}{g^2}F\bigg\vert_{N=0} = -\frac{16\pi^2}{g^2} + 3
\left[\frac{3 E^4}{8 \pi^2 g^2 v^4}\right]^{1/3}
+ O(E^2/v^2)\;.
\end{eqnarray}
These coincide with the results of Refs.~\cite{Khlebnikov:1991ue,Khlebnikov:1991th}.

\begin{figure}
  \centerline{\includegraphics[width=0.8\textwidth]{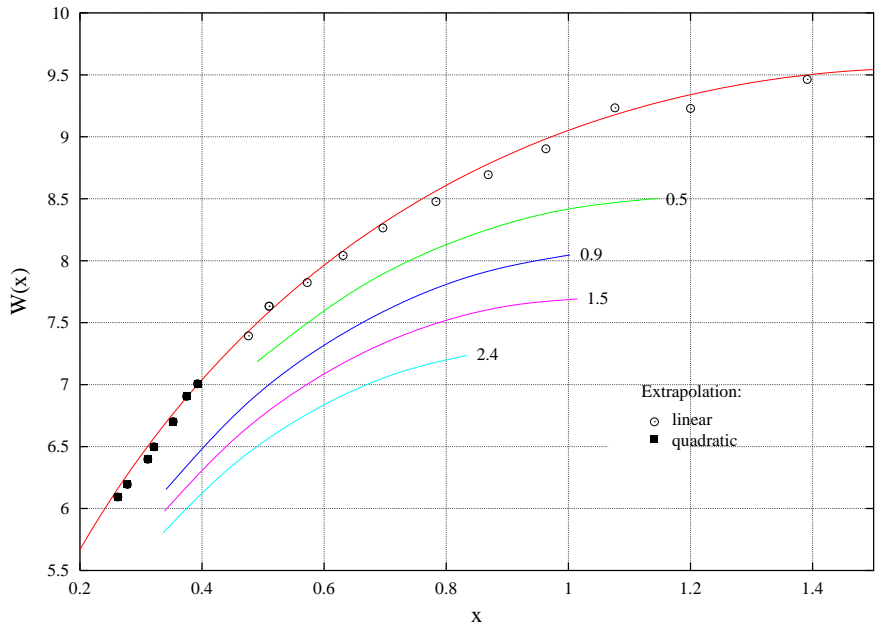}}
  \caption{Universal function $W(x)$ (the upper curve),
rescaled numerical results (solutions of the $T-\theta$ boundary value problem)
$(F+16\pi^2/g^2)/f$
at different energies (numbers near the graphs represent the values 
of the quantity $Eg\sqrt{2}/(4 \pi v)$), and the results of their extrapolation to 
zero $E$. Quantities $f$ and $x$ are determined by 
eqs.~\eqref{3-17},~\eqref{3-23,5}, $f\propto E^{4/3}$, 
$x\propto N/E^{4/3}$.}
\label{fig5}
\end{figure}
(iv) The $T/\theta$ boundary value problem summarized in Section 2 may be 
solved numerically. In Refs.~\cite{Bezrukov:2001dg,Bezrukov:2003dg},
the numerical study has been performed in the SU(2) Higgs model, so
we can directly check the numerical data of 
Refs.~\cite{Bezrukov:2001dg,Bezrukov:2003dg} against our analytical results.
Since the latter apply to relatively low energies only, this check 
involves extrapolation of the numerical results to $E\to 0$. To this end,
let us notice that according to eqs.~\eqref{3-14},~\eqref{3-15} the quantity
$$
W\equiv(F+16\pi^2/g^2)/f
$$
depends only on the combination $x\propto N/E^{4/3}$, up to corrections
$O(g^2E^2/v^2)$.In figure~\ref{fig5} we have plotted functions
$W(x)$ extracted from numerical data of 
Refs.~\cite{Bezrukov:2001dg,Bezrukov:2003dg} for different values of 
energy, and our analytical low energy prediction.
We see that though for energies close to the sphaleron energy
($Eg\sqrt{2}/(4\pi v) = 2.4$) the deviation between numerical and analytical results 
is fairly large, 
the extrapolation of the numerical results to zero energy
agrees with our analytical result.

\paragraph*{Acknowledgements}
The authors are indebted to V.~Rubakov for numerous
valuable discussions and criticism. This
research was supported in part by Russian Foundation for Basic Research grant
02-02-17398 and by  U.S.  Civilian Research and Development Foundation for
Independent States of FSU~(CRDF) award RP1-2364-MO-02.

\bibliographystyle{h-physrev4}
\bibliography{slac}

\begin{thebibliography}{10}

\bibitem{'tHooft:1976fv}
G.~'t~Hooft,
\newblock Phys. Rev. {\bf D14}, 3432 (1976).

\bibitem{Belavin:1975fg}
A.~A. Belavin, A.~M. Polyakov, A.~S. Shvarts and Y.~S. Tyupkin,
\newblock Phys. Lett. {\bf B59}, 85 (1975).

\bibitem{Ringwald:1990ee}
A.~Ringwald,
\newblock Nucl. Phys. {\bf B330}, 1 (1990).

\bibitem{Espinosa:1990qn}
O.~Espinosa,
\newblock Nucl. Phys. {\bf B343}, 310 (1990).

\bibitem{Khlebnikov:1991ue}
S.~Y. Khlebnikov, V.~A. Rubakov and P.~G. Tinyakov,
\newblock Nucl. Phys. {\bf B350}, 441 (1991).

\bibitem{Mueller:1991fa}
A.~H. Mueller,
\newblock Nucl. Phys. {\bf B364}, 109 (1991).

\bibitem{Khoze:1991sa}
V.~V. Khoze and A.~Ringwald,
\newblock CERN-TH-6082-91.

\bibitem{Arnold:1991pe}
P.~B. Arnold and M.~P. Mattis,
\newblock Phys. Rev. Lett. {\bf 66}, 13 (1991).

\bibitem{Mattis:1992bj}
M.~P. Mattis,
\newblock Phys. Rept. {\bf 214}, 159 (1992).

\bibitem{Tinyakov:1993dr}
P.~G. Tinyakov,
\newblock Int. J. Mod. Phys. {\bf A8}, 1823 (1993).

\bibitem{Rubakov:1992ec}
V.~A. Rubakov, D.~T. Son and P.~G. Tinyakov,
\newblock Phys. Lett. {\bf B287}, 342 (1992).

\bibitem{Tinyakov:1992fn}
P.~G. Tinyakov,
\newblock Phys. Lett. {\bf B284}, 410 (1992).

\bibitem{Mueller:1993sc}
A.~H. Mueller,
\newblock Nucl. Phys. {\bf B401}, 93 (1993).

\bibitem{Bonini:1999kj}
G.~F. Bonini, A.~G. Cohen, C.~Rebbi and V.~A. Rubakov,
\newblock Phys. Rev. {\bf D60}, 076004 (1999), [hep-ph/9901226].

\bibitem{Zakharov:1992dj}
V.~I. Zakharov,
\newblock Nucl. Phys. {\bf B371}, 637 (1992).

\bibitem{Porrati:1990rk}
M.~Porrati,
\newblock Nucl. Phys. {\bf B347}, 371 (1990).

\bibitem{Kuznetsov:1996cm}
A.~N. Kuznetsov and P.~G. Tinyakov,
\newblock Mod. Phys. Lett. {\bf A11}, 479 (1996), [hep-ph/9510310].

\bibitem{Kuznetsov:1997sf}
A.~N. Kuznetsov and P.~G. Tinyakov,
\newblock Phys. Lett. {\bf B406}, 76 (1997), [hep-ph/9704242].

\bibitem{Bezrukov:2003yf}
F.~Bezrukov and D.~Levkov,
\newblock quant-ph/0301022.

\bibitem{Bezrukov:2001dg}
F.~Bezrukov, C.~Rebbi, V.~Rubakov and P.~Tinyakov,
\newblock hep-ph/0110109.

\bibitem{Bezrukov:2003dg}
F.~Bezrukov {\em et~al.},
\newblock In preparation.  (2003).

\bibitem{Khlebnikov:1991th}
S.~Y. Khlebnikov, V.~A. Rubakov and P.~G. Tinyakov,
\newblock Nucl. Phys. {\bf B367}, 334 (1991).

\bibitem{Affleck:1981mp}
I.~Affleck,
\newblock Nucl. Phys. {\bf B191}, 429 (1981).

\bibitem{Forster:1977ip}
D.~Forster,
\newblock Phys. Lett. {\bf B66}, 279 (1977).

\bibitem{Callan:1978gz}
J.~Callan, Curtis~G., R.~F. Dashen and D.~J. Gross,
\newblock Phys. Rev. {\bf D17}, 2717 (1978).

\bibitem{Diakonov:1984hh}
D.~Diakonov and V.~Y. Petrov,
\newblock Nucl. Phys. {\bf B245}, 259 (1984).

\end{thebibliography}

\end{document}